\documentclass[letterpaper]{article} 
\usepackage[submission]{aaai2026}  
\usepackage{times}  
\usepackage{helvet}  
\usepackage{courier}  
\usepackage[hyphens]{url}  
\usepackage{graphicx} 
\usepackage{natbib}  
\usepackage{caption} 
\usepackage{algorithm}
\usepackage{algorithmic}
\usepackage{hyperref}
\usepackage[firstpage=true]{background}

\backgroundsetup{
  position=current page.south,
  angle=0,
  nodeanchor=center,
  vshift=5.5mm,
  opacity=0.5,
  scale=2,
  contents=[Presented at the RAI Workshop @ AAAI 2026]
}

\NewDocumentCommand{\fullref}{sm}{%
  \IfBooleanTF{#1}{%
    \namecref{#2} \nameref*{#2}%
  }{%
    \namecref{#2} \nameref{#2}%
  }%
}

\usepackage{newfloat}
\usepackage{listings}
\DeclareCaptionStyle{ruled}{labelfont=normalfont,labelsep=colon,strut=off} 
\floatstyle{ruled}
\newfloat{listing}{tb}{lst}{}
\floatname{listing}{Listing}
\title{Automated Reproducibility Has a Problem Statement Problem}
\author {
    Thijs Snelleman\textsuperscript{\rm 1},
    Peter Lundestad Lawrence\textsuperscript{\rm 2},
    Holger H. Hoos\textsuperscript{\rm 1,3} 
    Odd Erik Gundersen\textsuperscript{\rm 2}
}
\affiliations {
    \textsuperscript{\rm 1}Chair of AI Methodology, RWTH Aachen University, Aachen, Germany\\
    \textsuperscript{\rm 2}Department of Computer Science, Norwegian University of Science and Technology, Trondheim, Norway\\
    \textsuperscript{\rm 3} Leiden Institute of Advanced Computer Science, Leiden, The Netherlands\\
    snelleman@aim.rwth-aachen.de, peter.l.lawrence@ntnu.no\\
}

\usepackage{bibentry}

\begin{document}

\maketitle

\begin{abstract}

\textbf{Background.} Reproducibility is essential to the scientific method, but reproduction is often a laborious task. Recent works have attempted to automate this process and relieve researchers of this workload. However, due to varying definitions of reproducibility, a clear problem statement is missing.

\textbf{Objectives.}
Create a generalisable problem statement, applicable to any empirical study. We hypothesise that we can represent any empirical study using a structure based on the scientific method and that this representation can be automatically extracted from any publication, and captures the essence of the study.

\textbf{Methods.}
We apply our definition of reproducibility as a problem statement for the automatisation of reproducibility by automatically extracting the hypotheses, experiments and interpretations of $20$ studies and assess the quality based on assessments by the original authors of each study.

\textbf{Results.}
We create a dataset representing the reproducibility problem, consisting of the representation of $20$ studies. The majority of author feedback is positive, for all parts of the representation. In a few cases, our method failed to capture all elements of the study. We also find room for improvement at capturing specific details, such as results of experiments.

\textbf{Conclusions.}
We conclude that our formulation of the problem is able to capture the concept of reproducibility in empirical AI studies across a wide range of subfields. Authors of original publications generally agree that the produced structure is representative of their work; we believe improvements can be achieved by applying our findings to create a more structured and fine-grained output in future work.

\end{abstract}

\section{Introduction}

Reproducibility is widely considered a cornerstone of the scientific method~\cite{fidler2025reproducibility}, and although it is generally agreed that independent reproduction of published studies is indispensable for the advancement of science, such reproductions require substantial time investments of independent investigators~\cite{raff2019quantifying, Gundersen_Cappelen_Mølnå_Nilsen_2025}. Independent replications are necessary, yet often lead to less rewarding publications if published at all, due to their lack of novelty. In order to relieve independent investigators of the workload, the automatisation of reproducing studies to any extent would have a substantial impact. This automatisation has been attempted before, with various metrics to measure success: \citet{starace_paperbench_2025} introduced an `average replication score` based on `handcrafted' rubrics in coordination with the original authors to quantify reproduciblity, but their work lacks generalisability and scalability; \citet{hu2025repro} used the SSRP metric\footnote{\href{https://www.socialsciencereproduction.org/metrics}{https://www.socialsciencereproduction.org/metrics}} and a fine grained scoring structure by \citet{brodeur2024mass}, assessing how accurately their system is able to reproduce previous work using these metrics, as well as an applicability rate, which assesses to which degree the reproduced output is consistent with the original work. Although the metrics and rubric scores from \cite{starace_paperbench_2025, brodeur2024mass} may yield valid measurements, we believe these only capture parts or symptoms of the actual underlying problem of reproducibility, and lack a formal problem definition. Furthermore, the comparability of independent studies towards automating reproducibility is severely limited, due to the variation of metrics required to measure these diverging problem formulations.


In this study, we propose an approach that \emph{generalises} across studies; the approach can create representations without \emph{author intervention} and refrains from any instance-dependant rubrics, and can thus enable true automatisation. We formalise the problem statement based on terminology and existing structures of the scientific method~\cite{popper1934logic}. Our contributions are as follows;

\begin{itemize}
    \item Formal problem definition that generalises across empirical studies in AI, and thus can be used to create generalised metrics; in contrast to previous works which relied on instance-dependant rubrics.
    \item A proof-of-concept method that allows to automatically extract the problem statement of reproducibility for any empirical AI study.
    \item A dataset containing an empirical evaluation and corrected results of our automated method using $20$ published papers, which were evaluated by the authors of each study for our analysis.
\end{itemize}

\section{Related Work}
\label{sec:related_work}



Recently, several efforts have sought to automate the reproduction of scientific research; \citet{russo_advantages_2016} introduced ``executable'' papers, whereas \citet{brandmaier_automated_2025} suggested tools and frameworks for integrating reproducibility directly in the code repository when developing an experiment~\cite{gavish_universal_2011, jimenez_popperci_2017}. 
While these contributions are important steps towards automation, Large Language Models (LLM) offer the possibility to automatically reproduce scientific result, even when the authors do not 
apply specialised tools or frameworks to their publication. 

\citet{starace_paperbench_2025} evaluated multiple LLMs (o3-mini-high, GPT-4o, Gemini-2.0-Flash, DeepSeek-R1, o1-high and claude-3.5-sonnet) regarding their ability to reproduce scientific research. The best performing model reaches an \textit{average replication score} of $43.4\%$ ($\pm 0.8$). \citet{starace_paperbench_2025} emphasises the need for task decomposition. However, the evaluation of these decomposed tasks utilises various rubrics, based on the original authors' considerations of what constitutes reproducing their published work. This reduces the generalisability of the method to other studies, due the rubrics being defined per decomposed task per study. We also find that the method lacks a general definition of reproducibility; the rubrics are defined per paper, i.e., a per-instance definition which results in a separate rubric for each paper. Furthermore, the agents are not allowed to use code from the paper, which we consider to be a key element of documentation of a study. The evaluation of the reproducibility agents by \citet{starace_paperbench_2025} was carried out by an LLM judge, which achieves a performance of $0.84$ F1 score on a rather small evaluation set of twenty studies. This makes it difficult to determine whether this judge, and thus the 'average replication score', is sufficiently representative to apply to other automated reproducibility systems for evaluation.

\citet{hu2025repro} also aimed to leverage LLMs to automate the reproduction of studies, within social sciences, in a single-agent system. 
Their agent was provided with all the paper, data, pre-installed dependencies and detailed description of the task. The agent was set to determine a reproducibility score from 1 to 4 for each publication evaluated, where 1 corresponds to the least and 4 to the most reproducible work. Scores 1 and 4 are objective binary statements (true or false), resulting in a strict and objective top and bottom score; scores 2 and 3 require assumptions about what ``minor issues/inconsistencies'' entail, which is more vague and abstract, as they require interpretation. The ground truth reproducibility score was set by the authors.
The best-performing agent, REPRO-agent, achieved an accuracy of 36.6\% when its assigned scores were compared against the ground truth. 

Reproducibility scores in this benchmark are dependent on the availability of code and data in the evaluated publications; consequently, papers without code or data were excluded. REPRO-bench focuses on reproducibility in the social sciences, where reproducibility is often closely tied to data availability. However, this does not generalise across all scientific disciplines. For example, a computer science study comparing two search algorithms, $A$ and $B$, might claim that $A$ consistently outperforms $B$ under specified conditions. Such a claim could be reproducible without relying on the original dataset. Therefore, a reproducibility metric that depends strictly on the presence of code and data may not be fully applicable to computer science research.
\citet{xiang_scireplicate-bench_2025} implemented a multiagent system for reproducing scientific research consisting of two parts; a paper agent and code agent. The paper agent extracts information from the paper and creates a literature report. The code agent uses this report as input, searching through any code and files, as well as conducting a web-search for relevant information. The code agent compiles and runs the resulting code, and is able to respond to feedback from the compiler, allowing it to troubleshoot and improve the implementation. 
\citet{xiang_scireplicate-bench_2025} evaluated the paper agent and the code agent individually. The evaluation was done using CodeBLEU with ground-truth code, a novel reasoning accuracy graph, execution accuracy and recall for intra-file dependencies, cross-file dependencies and external APIs. The results showed overall that the agents perform better at summarising the algorithms and code, but lag behind in terms of implementation and execution.

Similar efforts have been made by \citet{zhao_autoreproduce_2025}, using a researcher agent and a coding agent. The researcher agent makes use of a paper lineage algorithm, in order to determine the most relevant citations, thus gaining further knowledge about the problem and domain. In addition, the researcher agent tries to extract the method and experiment from the paper. The authors evaluated the agents ability to `understand' the paper, the code and the execution of the experiment. They concluded that the implementation and execution of code is a difficult task for the agent based on the large performance gaps. As seen before, the authors introduced their own metrics (Align-Score and Exec-Score) to evaluate their system, which makes comparability to other automation methods difficult.

Common across all the previously mentioned studies~\cite{starace_paperbench_2025, hu2025repro, xiang_scireplicate-bench_2025, zhao_autoreproduce_2025} is the limited attention to the underlying relation between the reproducibility of a study and the scientific method. Decomposing tasks from a formal problem statement to reproduce the research using the scientific method as framework is essential for the comparability and objective evaluation of such systems. In addition, the papers mentioned all utilise a single or dual agent system. Thus, they leave a gap in terms of solving the problem of automatic reproducibility using larger multiagent systems. Recent advancements have been made into such multiagent systems~\cite{chen_autoagents_2024}. Complex problems are often solved by multiple actors, concurrently working on sub-tasks~\cite{ozturk_multiagent_2010}. Thus, the agents produce a solution to the problem through collaboration, where each agent's attention is on a smaller task towards solving the problem. We believe that, through our problem formulation, we can create such an agent that automatically extracts the problem from any study and provide this as input to the other agents in a structured and easily distributable tasks.

\section{Background}
To reduce the issues discussed in the previous section, and to provide a generalisable framework applicable to all empirical AI studies, we aim to reframe the formulation of the problem to allow for comparability of outcomes between various automated reproducibility agents.
We consider the following notion of reproducibility, based on \citet{gundersen2021fundamental}: 
Based on the documentation provided by the original authors, independent investigators are able to conduct similar experiments, the outcomes of which can be analysed and interpreted to support the hypotheses of the original investigators. 

\begin{figure}
    \centering
    \includegraphics[width=0.9\linewidth]{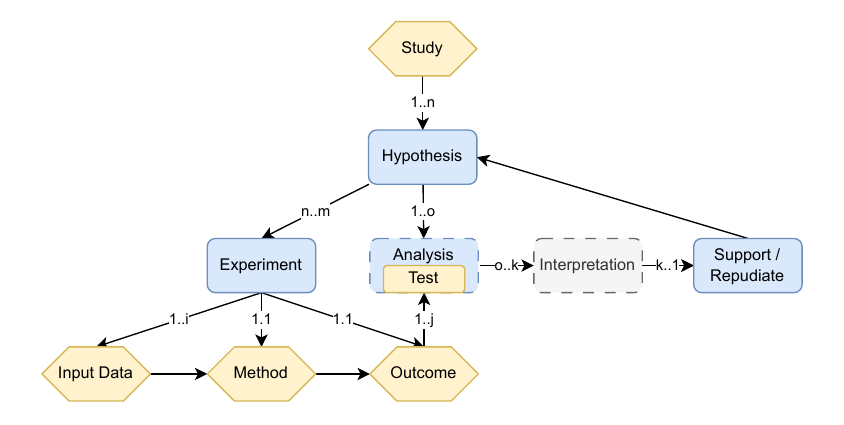}
    \caption{A general overview of our problem statement of reproducing an empirical study. We model the problem as a graph: A study contains one or more hypotheses, evaluated and tested through multiple experiments. Outcomes are analysed and interpreted to support or repudiate a given hypothesis. The analysis is reduced to elements needed for assessing the outcome of experiments. The interpretation element is graphically distinguished, since we treat it as static, whereas traditionally, these can be more flexible.}
    \label{fig:scientific_pipeline}
\end{figure}

From this, we derive our problem statement, formulated in terms of the scientific method and summarised in \autoref{fig:scientific_pipeline}. 
We consider studies to contain hypotheses, which are linked to experiments. 
Each experiment contains one or more sets of input data (e.g.{} data sets) and applies some method or strategy to produce outcomes (e.g. measurements). 
These outcomes are then analysed using, for example, statistics and calculated metrics as well as some form of testing (e.g. statistical testing, direct comparison of values or visual representation); the results of these analyses are then interpreted to support the hypotheses. 
As shown in the diagram, we opt for a flexible representation, where each experiment can be linked with multiple hypotheses, outcomes can be subject to multiple analyses, and interpretations can be based on various analyses over multiple experiments. 

In the context of our work, some practical considerations arise: 
Firstly, for manual reproduction by human investigators, the interpretation of outcomes may change, but still yield support for the hypotheses and thus successful reproduction; in an automated setting, we find this to be a liability, and thus treat these interpretations relatively static. Secondly, we generally simplify the analysis to the extraction of `results', and we describe the task as the extraction of values based on the determined metrics and statistics. Using this structure, automated reproducibility can be achieved and measured based on producing similar results, which pass the (statistical) test defined by the authors, thus supporting the same interpretations and upholding the hypotheses defined by the authors. This allows for broad adaptation and generalisability; 
in particular, the capability of any system to reproduce an empirical study can be measured by determining what part of the graph can be reproduced to uphold the hypotheses stated by the authors.

\section{Method}
\label{sec:method}
To apply our problem statement practically, we consider the first step towards automating reproducibility to be able to automatically extract all elements from the diagram in \autoref{fig:scientific_pipeline} from any publication. 
We constructed a relatively simple prompt and presented this, together with the PDF of each publication, to Google Gemini 2.5 Pro~\cite{comanici2025gemini}\footnote{We used a  temperature setting of $t = 0.0$ to reduce model stochasticity}. We then reviewed the output of the LLM with the first author of each work, to assess the quality of the LLM-based analysis. 
The authors were asked to correct any mistakes made by the LLM in its phrasings, links between hypotheses, experiments and interpretations, as well as experiment details, such as measured outcomes, applied statistics, strategies and how tests are used for assessing outcomes. 
For each element in the problem statement, the authors were asked to rate the overall answer of the LLM on a 5-point, and one 7-point, Likert scale. The full prompt, outputs, review form and outcomes can be found in our GitHub repository\footnote{\href{https://github.com/thijssnelleman/automated-reproducibility}{https://github.com/thijssnelleman/automated-reproducibility}}. 
The authors of each article were informed about the formulation of the problem and the objective of the extraction to ensure representative evaluations of the LLM output.


During the development of our methodology, we noted one crucial difference between our set-up and the real-world setting; it is rather uncommon for empirical AI studies to explicitly formulate hypotheses. Rather, authors generally state research questions and findings, instead of stating a hypothesis with an expected outcome. Despite this, we still find value in our method using a slight adaptation; the hypothesis that is constructed from the latent representation of the study should be considered a \emph{post-hoc} hypothesis. From the perspective of independent reproduction, the expected outcome of the experiment is to draw the same conclusions as the original investigators.

We applied few-shot prompting to obtain our results. The LLM was given various examples on how to determine the answer and structure its output; we provided the LLM with hints, such as sections that may contain the target information, as well as possible keywords that may signal essential information. 
Furthermore, we queried the model multiple times for three candidate publications~\cite{dettmer2024weighted, Gundersen_Cappelen_Mølnå_Nilsen_2025, snelleman2024edge}, to improve the quality of the prompt and subsequent output. This should not be interpreted as few-shot learning, as the model was not presented with any feedback; the author feedback was only used to improve the quality of the prompt. Afterwards, we prompted the model to produce the hypotheses, experiments and interpretations of outcomes for the $20$ publications listed in \autoref{tab:paper_table}.

\begin{table}
    \centering
    \begin{tabular}{lr}
         Paper & \# Tokens \\
         \hline
         \citet{anastacio2022instance} & 3 355\\
         \citet{benjamins2025carpsframeworkcomparingn} & 9 031\\
         \citet{BerEtAl25} & 7 225\\
         \citet{BosEtAl25} & 10 579\\
         \citet{dettmer2024weighted} & 4 129\\
         \citet{DieEtAl24} & 6 709\\
         \citet{downing2023evolution} & 6 967\\
         \citet{eimer2023hyperparameters} & 11 869\\
         \citet{fehring2025growing} & 1 291\\
         \citet{fleten2024applying} & 6 193\\
         \citet{jankovic2022trajectory} & 2 065\\
         \citet{jekic2025examining} & 8 257\\
         \citet{KauEtAl25} & 5 935\\
         \citet{ParEtAl25} & 3 613\\
         \citet{renting2025towards} & 3 097\\
         \citet{SkaEtAl25} & 11 095\\
         \citet{ShaHoo24} & 6 709\\
         \citet{snelleman2024edge} & 3 871\\
         \citet{toussaint2025edc} & 3 871\\
         \citet{WasEtAl25} & 3 355\\
    \end{tabular}
    \caption{All publications used for the evaluation of our method, sorted alphabetically by first author.}
    \label{tab:paper_table}
\end{table}

\section{Empirical Evaluation}

\begin{figure}
    \centering
    \includegraphics[width=\linewidth]{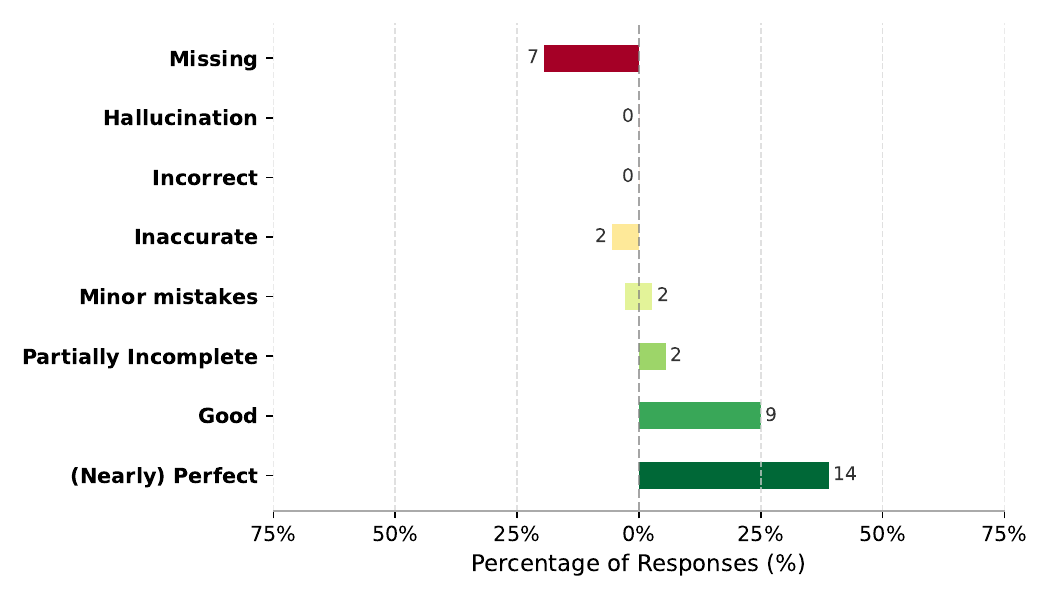}
    \caption{Evaluation of the hypotheses captured by the LLM by the original authors, using a 7-point Likert scale, including missing hypotheses supplemented by the authors.}
    \label{fig:hypothesis_likert}
\end{figure}

\begin{figure}
    \centering
    \includegraphics[width=\linewidth]{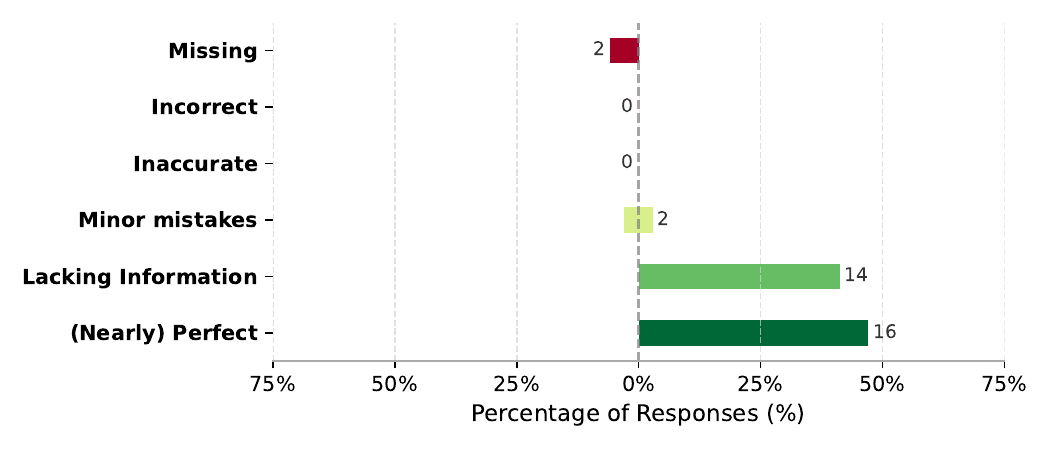}
    \caption{Evaluation of the extracted experiment descriptions by the original authors, using a 5-point Likert scale plot. This includes missing experiments supplemented by the authors.}
    \label{fig:experiment_description_likert}
\end{figure}

\begin{figure}
    \centering
    \includegraphics[width=\linewidth]{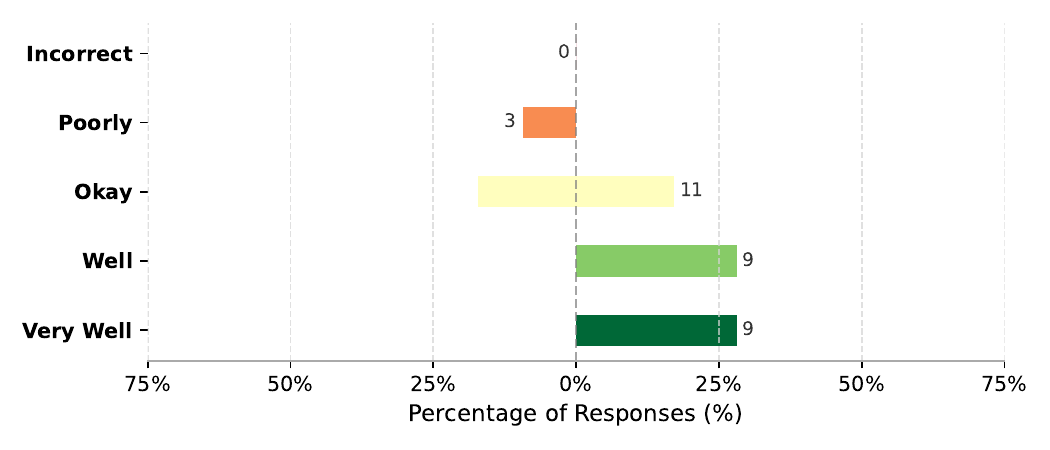}
    \caption{Evaluation of the extracted experiment details by the original authors, using a 5-point Likert scale.}
    \label{fig:experiment_details_likert}
\end{figure}

\begin{figure}
    \centering
    \includegraphics[width=\linewidth]{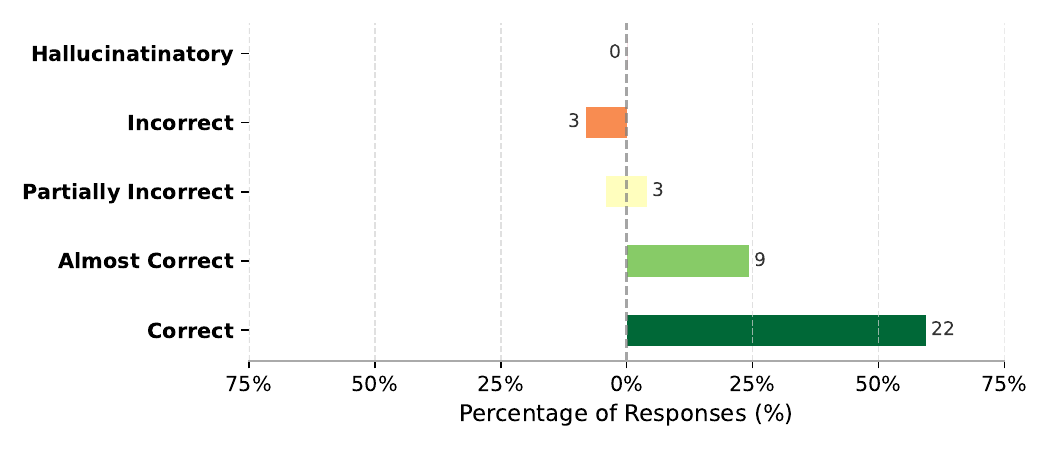}
    \caption{Evaluation of the extracted experiment interpretations by the original authors, using a 5-point Likert scale. Authors were given the opportunity to adapt the phrasing.}
    \label{fig:interpretation_likert}
\end{figure}

For each paper, the authors were asked to rate the output of our procedure on a 5-point or 7-point Likert scale and to correct possible mistakes. 
The ratings for each element of our analysis (see \autoref{fig:scientific_pipeline}) are shown in \autoref{fig:hypothesis_likert}, \autoref{fig:experiment_description_likert}, \autoref{fig:experiment_details_likert} and \autoref{fig:interpretation_likert}. We have summarised the error rates of our approach in \autoref{tab:pipeline_mistakes}. We found that in $75.00\%$ of the studies, our method was able to correctly capture all elements; in these cases, all hypotheses and experiments were represented at least to some degree. Based on the Likert scale results, it is apparent that our method was assessed rather positively by the authors; overall, it appears to be able to capture the hypotheses, experiment descriptions and details, as well as the interpretation of outcomes quite well. 

However, there are some noteworthy caveats to this assessment. In \autoref{fig:hypothesis_likert}, we see that in six cases, the methodology was not able to capture the hypotheses of a given study. 
Upon closer inspection, we found that in all cases, the method was able to capture one or more hypotheses correctly, but failed to determine the full set of hypotheses. 
In one such study, \citet{BosEtAl25}, the authors investigated \emph{nine} hypotheses in total, of which our method captured seven. 
In another case, \citet{benjamins2025carpsframeworkcomparingn}, the method was only able to determine one out of two hypothesis. 
When comparing the token counts of these two publications against those of the remainder of our data set, as seen in \autoref{tab:paper_table}, we observe that they are substantially larger, and thus the failure of capturing all details could possibly be attributed to the demands placed upon the LLM in terms of context length. 

Similarly, we noticed in \autoref{fig:experiment_description_likert} that our method receives positive evaluations, but has failed in two cases to capture an entire experiment, which occurred in \citet{benjamins2025carpsframeworkcomparingn} and \citet{BerEtAl25}. The latter also has a relatively large number of tokens. However, in the case of \citet{SkaEtAl25} and \citet{eimer2023hyperparameters}, no missing experiments were observed, even though these are the largest studies in the dataset in terms of token count; this indicates that token count alone does not explain the difficulties encountered with some studies. 

We note that for the study by \citet{eimer2023hyperparameters}, the first author stated that our approach merged three experiments into one; considering the problem statement from \autoref{fig:scientific_pipeline}, this author regarded this as a minor issue. 
We further note that, although the majority of extracted hypotheses were evaluated positively, in \autoref{tab:pipeline_mistakes}, we can see that in $65.52\%$ of the cases the authors wished to adapt the hypothesis extracted by the LLM to reflect the work more closely. However, it can also be seen in \autoref{tab:pipeline_mistakes} that on average, $43$ characters were changed by the authors, which corresponds to only $14.90\%$ of the statement on average, showing that, although authors wished to adapt the captured hypothesis, changes were relatively minor in terms of textual changes. Note that the Levenshtein distance does not cover any semantic changes of the statements.

\begin{table}
    \centering
    \begin{tabular}{l|rr}
         & Error \# & Proportion \\
        Hypothesis Statements & 19 & 65.52 \% \\
        Hypothesis Edit Distance & 43 & 14.90 \% \\
        Interpretation Statements & 9 & 24.32 \% \\
        Interpretation Edit Distance & 35 & 4.79 \% \\
        \hline
        Experiment Hypothesis links & 6 & 18.75 \% \\
        Interpretation Hypothesis links & 0 & 0.00 \% \\
        Interpretation Experiment links & 2 & 5.41 \% \\
        \hline
        Experiment Metrics & 15 & 46.88 \% \\
        Experiment Statistics & 9 & 28.12 \% \\
        Experiment Strategy & 10 & 31.25 \% \\
        Experiment Results & 1103 & 69.63 \% \\

    \end{tabular}
    \caption{Error rates of our method on phrasing hypotheses and interpretations, extracting links in the problem statement, experiment details and results. The edit distance, i.e.{} the amount of corrected characters in a statement, was calculated using the Levenshtein distance~\cite{miller2009levenshtein} and rounded up for averages. The error count in the experiment results includes missing values as well as incorrect values.}
    \label{tab:pipeline_mistakes}
\end{table}

In \autoref{fig:experiment_details_likert} the authors evaluation of the ability of the method to capture the details of each experiment. Overall the authors evaluate the output quite positively, but this is somewhat conflicting with the results in \autoref{tab:pipeline_mistakes}, where we can see for example that $69.63\%$ of the experiment results were either corrected or not fully captured, as well as the experiment metrics needing to be corrected in $46.88\%$ of the cases. However, the authors found overall that the general spirit and goal of the experiment was captured, albeit with a substantial amount of mistakes when capturing outcome values for example. Another important note with regards to capturing experiment results, is visualised outcomes; it is not uncommon in an empirical study to visualise certain parts of the experiments with for example box-plots, line graphs or histograms, and interpret the outcome visually. Although the LLM made attempts to extract this information from the paper, the results were quite unstable. The results often defaulted to extracting this information from the text rather than from the image, especially when the images were not vectorised or rasterised within the PDF.

The interpretation of the results were received quite positive as well, as seen in \autoref{fig:interpretation_likert}, and also needed substantially less adaptations compared to the hypotheses as seen in \autoref{tab:pipeline_mistakes}; $24.32\%$ of the interpretations were edited, with an average change of $4.79\%$ per statement. Overall, we noticed that the amount of quoting and paraphrasing of the original paper was much more substantial than in the hypotheses, thus likely to play a major role in reducing the amount of mistakes made by the LLM. 

\section{Discussion}


One of the most challenging issues for our automated extraction method consists of dealing with visual depictions of results, such as graphs and diagrams.
Overall, we observed that our method is capable of extracting structured results, such as tables, with relative ease, and with an improved prompt, it should be possible to reach even higher accuracy in this type of analysis. We believe that the difficulty of dealing with figures can, in principle, be addressed -- for example, by providing clear instructions to the LLM on how to capture and interpret the content of figures rather than focussing merely on the respective descriptions provided in the text of a given publication. 
Still, the multi-modality of the data is likely to remain challenging.

As mentioned previously, in a few cases, we also found that our method was not able to extract all hypotheses and experiments.
On the one hand, this indicates that our prompt is unable to generalise properly to all empirical AI studies; we believe that improvements in this regards are possible, using our dataset as training data. On the other hand, it also indicates that in some cases, a clear phrasing of hypotheses or research questions within a given study is essential for capturing the essence of the work -- be it by human readers or automated methods. 

Finally, we observed a potential link between the number of tokens, i.e.{} the length of a given study, and the accuracy of the results obtained from our LLM-based approach. This suggests that that extracting latent representations of hypotheses becomes more complicated for longer, more complex publications.

\section{Conclusion \& Future Work}

In this work, we aimed to phrase a problem statement for reproducibility to enable solutions that can generalise to any empirical AI paper. We designed a problem representation that is based on the foundations of the scientific method. We find that is able to capture the essential elements of any empirical AI study and believe it could be generalised beyond our field as well. Furthermore, the underlying unified graph structure allows independent studies to measure similar metrics based on what elements of the graph they were able to reproduce, for example by counting how many hypothesis interpretations were upheld by their automated reproducibility method. Thus, our method can serve both as a structured input problem, as well as a comparable output structure across independent studies.

We applied our representation to automatically extract the problem from any PDF and reviewed its capabilities on $20$ studies in consultation with the respective first authors. Overall, we found that our methodology is capable of capturing the hypotheses, experiments and outcome interpretations of these studies. However, in some cases, our method failed to capture all essential information required, e.g., due to missing hypotheses and experiments or details of experiments which should be improved upon through for example extensive prompt engineering or even post-training of LLMs on this task.

Our work serves as  a proof of concept, to enable other methods, such as \citet{starace_paperbench_2025} and \citet{bhaskar_reproscreener_2024}, to solve the problem of reproducibility through a generalisable framework, that in turn enables clear optimisation goals to measure improvements, which were found highly necessary in both works. In future work, we believe it necessary to improve on our automated extraction method; the set-up used here is rather simplistic, and improvement is possibly achievable by using our published dataset, allowing for more accessible and detailed data structures for any automated reproducibility system to solve.

\section{Acknowledgments}

We would like to thank our colleagues, specifically Kevin Coakley, whose expertise in prompt engineering has substantially accelerated the application side of this project. We would also like to thank Justin Dettmer, for his support and feedback during the early stages of this project. Finally, we would like to thank all the authors who participated in reviewing the results produced by our method; their time and input was essential to our work.

This project was partially funded by an Alexander von Humboldt-Professorship in AI held by Holger H. Hoos, and by the Norwegian Research Council under the Robust Intelligent Control project (RICO, 329730).

\bibliography{aaai2026}

\makeatletter
\@ifundefined{isChecklistMainFile}{
  \newif\ifreproStandalone
  \reproStandalonetrue
}{
  \newif\ifreproStandalone
  \reproStandalonefalse
}
\makeatother

\ifreproStandalone
\documentclass[letterpaper]{article}
\usepackage[submission]{aaai2026}
\setlength{\pdfpagewidth}{8.5in}
\setlength{\pdfpageheight}{11in}
\usepackage{times}
\usepackage{helvet}
\usepackage{courier}
\usepackage{xcolor}
\frenchspacing

\begin{document}
\fi
\setlength{\leftmargini}{20pt}
\makeatletter\def\@listi{\leftmargin\leftmargini \topsep .5em \parsep .5em \itemsep .5em}
\def\@listii{\leftmargin\leftmarginii \labelwidth\leftmarginii \advance\labelwidth-\labelsep \topsep .4em \parsep .4em \itemsep .4em}
\def\@listiii{\leftmargin\leftmarginiii \labelwidth\leftmarginiii \advance\labelwidth-\labelsep \topsep .4em \parsep .4em \itemsep .4em}\makeatother

\setcounter{secnumdepth}{0}
\renewcommand\thesubsection{\arabic{subsection}}
\renewcommand\labelenumi{\thesubsection.\arabic{enumi}}

\newcounter{checksubsection}
\newcounter{checkitem}[checksubsection]

\newcommand{\checksubsection}[1]{%
  \refstepcounter{checksubsection}%
  \paragraph{\arabic{checksubsection}. #1}%
  \setcounter{checkitem}{0}%
}

\newcommand{\checkitem}{%
  \refstepcounter{checkitem}%
  \item[\arabic{checksubsection}.\arabic{checkitem}.]%
}
\newcommand{\question}[2]{\normalcolor\checkitem #1 #2 \color{blue}}
\newcommand{\ifyespoints}[1]{\makebox[0pt][l]{\hspace{-15pt}\normalcolor #1}}

\section*{Reproducibility Checklist}

\vspace{1em}
\hrule
\vspace{1em}


\checksubsection{General Paper Structure}
\begin{itemize}

\question{Includes a conceptual outline and/or pseudocode description of AI methods introduced}{(yes/partial/no/NA)}
Yes

\question{Clearly delineates statements that are opinions, hypothesis, and speculation from objective facts and results}{(yes/no)}
Yes

\question{Provides well-marked pedagogical references for less-familiar readers to gain background necessary to replicate the paper}{(yes/no)}
Yes

\end{itemize}
\checksubsection{Theoretical Contributions}
\begin{itemize}

\question{Does this paper make theoretical contributions?}{(yes/no)}
No
\end{itemize}

\checksubsection{Dataset Usage}
\begin{itemize}

\question{Does this paper rely on one or more datasets?}{(yes/no)}
Yes

\ifyespoints{If yes, please address the following points:}
\begin{itemize}

	\question{A motivation is given for why the experiments are conducted on the selected datasets}{(yes/partial/no/NA)}
	NA

	\question{All novel datasets introduced in this paper are included in a data appendix}{(yes/partial/no/NA)}
	Yes

	\question{All novel datasets introduced in this paper will be made publicly available upon publication of the paper with a license that allows free usage for research purposes}{(yes/partial/no/NA)}
	Yes

	\question{All datasets drawn from the existing literature (potentially including authors' own previously published work) are accompanied by appropriate citations}{(yes/no/NA)}
	NA

	\question{All datasets drawn from the existing literature (potentially including authors' own previously published work) are publicly available}{(yes/partial/no/NA)}
	NA

	\question{All datasets that are not publicly available are described in detail, with explanation why publicly available alternatives are not scientifically satisficing}{(yes/partial/no/NA)}
	NA

\end{itemize}
\end{itemize}

\checksubsection{Computational Experiments}
\begin{itemize}

\question{Does this paper include computational experiments?}{(yes/no)}
Yes

\ifyespoints{If yes, please address the following points:}
\begin{itemize}

	\question{This paper states the number and range of values tried per (hyper-) parameter during development of the paper, along with the criterion used for selecting the final parameter setting}{(yes/partial/no/NA)}
	Yes

	\question{Any code required for pre-processing data is included in the appendix}{(yes/partial/no)}
	NA

	\question{All source code required for conducting and analyzing the experiments is included in a code appendix}{(yes/partial/no)}
	Yes

	\question{All source code required for conducting and analyzing the experiments will be made publicly available upon publication of the paper with a license that allows free usage for research purposes}{(yes/partial/no)}
	Yes
        
	\question{All source code implementing new methods have comments detailing the implementation, with references to the paper where each step comes from}{(yes/partial/no)}
	Yes

	\question{If an algorithm depends on randomness, then the method used for setting seeds is described in a way sufficient to allow replication of results}{(yes/partial/no/NA)}
	Yes

	\question{This paper specifies the computing infrastructure used for running experiments (hardware and software), including GPU/CPU models; amount of memory; operating system; names and versions of relevant software libraries and frameworks}{(yes/partial/no)}
	Partial

	\question{This paper formally describes evaluation metrics used and explains the motivation for choosing these metrics}{(yes/partial/no)}
	Yes

	\question{This paper states the number of algorithm runs used to compute each reported result}{(yes/no)}
	Yes

	\question{Analysis of experiments goes beyond single-dimensional summaries of performance (e.g., average; median) to include measures of variation, confidence, or other distributional information}{(yes/no)}
	No

	\question{The significance of any improvement or decrease in performance is judged using appropriate statistical tests (e.g., Wilcoxon signed-rank)}{(yes/partial/no)}
	No

	\question{This paper lists all final (hyper-)parameters used for each model/algorithm in the paper’s experiments}{(yes/partial/no/NA)}
	NA

\end{itemize}
\end{itemize}
\ifreproStandalone
\end{document}
\fi

\end{document}